\pdfoutput=1
\documentclass[11pt]{article}

\usepackage[final]{acl}

\usepackage{times}
\usepackage{latexsym}
\usepackage{graphicx}
\usepackage{comment}
\usepackage{parskip}
\usepackage{varwidth}

\usepackage[most]{tcolorbox}
\usepackage{xcolor}

\usepackage{csquotes}

\definecolor{promptcolor}{RGB}{0, 102, 204}
\definecolor{responsecolor}{RGB}{51, 153, 51}
\definecolor{BrickRed}{RGB}{143, 20, 2}

\newtcolorbox{promptresponsebox}{
    colback=white,
    colframe=black,
    arc=0mm,
    outer arc=0mm,
    boxrule=0.5pt,
    leftrule=10pt,
    toprule=0pt,
    bottomrule=0pt,
    rightrule=0pt,
    left=12pt,
    right=12pt,
    top=8pt,
    bottom=8pt,
    before skip=10pt,
    after skip=10pt,
    fontupper=\color{black},
    fonttitle=\bfseries,
}

\newtcolorbox{tagbox}{
  colback=gray!10!white,
  colframe=gray!80!black,
  boxrule=0.5mm,
  arc=2mm,
  left=1mm,
  right=1mm,
  top=1mm,
  bottom=1mm,
  boxsep=1mm,
  fontupper=\ttfamily,
  varwidth boxed title
}

\usepackage[T1]{fontenc}
\usepackage[utf8]{inputenc}
\usepackage{microtype}
\usepackage{inconsolata}

\title{Health Insurance Coverage Rule Interpretation Corpus: Law, Policy, and Medical Guidance for Health Insurance Coverage Understanding}

\author{Mike Gartner}

\newcommand{\BibTeX}{\rm B\kern-.05em{\sc i\kern-.025em b}\kern-.08em\TeX}

\begin{document}

\pagestyle{plain}

\maketitle 
\begin{abstract}
  U.S. health insurance is complex, and inadequate understanding and limited access to justice have dire implications for the most vulnerable. 
  Advances in natural language processing present an opportunity to support efficient, case-specific understanding, 
  and to improve access to justice and healthcare. Yet existing corpora lack context necessary for assessing even simple cases. 
  We collect and release a corpus of reputable legal and medical text related to U.S. health insurance. We also introduce an outcome prediction 
  task for health insurance appeals designed to support regulatory and patient self-help applications, and release a labeled benchmark for our task, 
  and models trained on it.
\end{abstract}

\section{Introduction}

      Health insurance coverage issues in the U.S. pose serious problems for patients. They cause delays, forgone care, and detrimental health outcomes. Sometimes patients receive services despite denials, incurring bills, debt, and collections lawsuits. For the most vulnerable, there are calamitous effects \cite{kffDenialSurvey, commonwealthDebtArticle, kffDebtArticle}. 
    
     Evaluating coverage rules requires precise understanding of a complex web of case-specific laws,  contracts, and medical literature. In cases where existing automation does not exist, fails, or is disputed, specialists such as medical coders, pharmacists, doctors, and attorneys do such work manually. The processes are time-consuming and error-prone, and successful navigation often involves tasks like synthesis of medical literature and contracts.
      
      Advances in natural language understanding (NLU) present an opportunity to improve patient outcomes. In particular, extractive and generative models can increase the efficiency with which specialists understand cases. Tools that support efficient and accurate coverage understanding stand to serve patients, caseworkers, and regulators alike. They can lower unnecessary administrative costs, and improve both health and financial outcomes.
      
      However, there are barriers preventing responsible and effective use of such tools. Most notably, there is a dearth of high-quality, curated corpora containing information sufficient to correctly answer coverage questions. 
      
      We make four contributions. First, we collect and release a corpus of text that informs understanding of health insurance coverage rules. Second, we present an `appeal adjudication' task.  Third, we curate and release an annotated dataset to serve as a benchmark for the task. Finally, we train, evaluate, and release baseline models using our benchmark.
      
      The datasets and code for all of our contributions have been made publicly available\footnote{Code was released at \href{https://github.com/TPAFS/hicric}{https://github.com/TPAFS/hicric}.} under a permissive license.
      This work was previously accepted and presented as a workshop paper at the AI for Access to Justice workshop at JURIX 2024.

\section{Acknowledgements}
  It is my pleasure to thank Leigh Thorne for her feedback on the interim annotation guidelines, and Jeanne Alicandro and
  anonymous referees for providing detailed comments and suggestions on this manuscript.

\section{Related Work}

\textbf{Pretraining Corpora.}
Domain adaptive pretraining can improve performance on downstream tasks \cite{dont-stop-pretraining, when-does-pretraining-help}. High quality pretraining corpora for both legal and medical applications  abound.

In the legal domain, there is a rich history of pretraining models via masked language objectives \cite{bertlawschool, legalbert, pileoflaw, multilegalpile, budget-longformer, legalllama}.

The medical domain shares a similar history \cite{clinicalbert, clinicalbert2, domain-specific-biomedical, meditron}.

Recent years have seen an increasing focus on using language models to support arbitrary downstream tasks (e.g. \cite{budget-longformer, flawn-t5}). For tasks involving complex domain specific context, high quality pretraining corpora are critical. 
This is especially true for legal and medical question answering.

Our corpus shares some overlap with two existing datasets: Pile of Law \cite{pileoflaw} and the Guidance dataset \cite{meditron}. 
The overlap can be understood in detail by consulting Appendix \ref{sec:detailedcomposition}.

To our knowledge, this work is among the first few efforts targeting curation of pretraining data in the intersection of legal and medical domains. 
The only other works of which we are aware are that of \cite{legalreelectra}, which focused on tasks involving personal injury cases,
and that of \cite{spanish_insurance_coverage} and \cite{health_system_scale_model} which involve health insurance coverage tasks.

\textbf{Supervised Finetuning Data.}
There has also been a large body of work focused on curating data for specific legal and medical tasks.

In the legal domain, people have developed datasets and models to support entity and clause selection \cite{cuad, maud}, multiple choice question answering \cite{lexglue, casehold}, span-based question answering \cite{squadv2, privacyQA, policyQA}, extractive summarization \cite{legal-extractive, mining-legal-arguments-in-court}, retrieval \cite{legal-case-matching}, and document and word level classification \cite{neural-legal-judgement-pred, ledgar}, among other tasks.

In the medical domain, people have developed datasets and models to support multiple choice question answering \cite{medmcqa}, span-based question answering \cite{medQA, pubmedQA}, document and clause classification \cite{clinicalbert, clinicalbert2}, and language modeling \cite{meditron, deepmindllm}, among other tasks.

In this work we construct a benchmark dataset for a document classification task. We also describe how to adapt our (pseudo)-annotation pipeline to yield related, but distinct benchmarks.

\textbf{Case Outcome Prediction Tasks.}
Our task shares qualities with existing legal judgement prediction tasks \cite{court-judgement-prediction, swiss-judgement, lextreme, legalbench}.

In particular, a few tasks presented in recent literature share many similarities.

For example, 
our task is at a high level the same as that presented in \cite{spanish_insurance_coverage}. The task there is to predict health insurance coverage outcomes from Spanish 
language clinical notes. Our task differs in a few critical ways. It is a three class problem, 
focused on English, and concerned with both expert and layperson descriptions.

Similarly, \cite{health_system_scale_model} introduces a general purpose medical language model, and a health insurance denial task. 
Like \cite{spanish_insurance_coverage}, it focuses on making predictions from clinical notes. The authors did not release model weights 
nor training data, so the results are not reproducible. Our work presents a distinct task, and a more 
open source approach. We have released the training data and code under permissive licenses.

Finally, we note that outcome prediction tasks in the literature are often actually retrospective outcome classification 
tasks \cite{legal_judgement_prediction_do_it_right}. The tasks, benchmarks, and models do not support real 
world forecasting. Instead, they `predict' case outcomes from text which is not and could not be produced before adjudication. 
Our focus is on real world forecasting applications, and our methods reflect that focus.

\section{A Corpus}

Our dataset consists of documents from diverse sources. It includes U.S. federal and state law, insurance contracts, official regulatory guidance, agency opinions and policy briefs, official coverage rules for Medicaid and Medicare, and summaries of appeal adjudications. In total, the dataset contains 8,311 documents, 419 million words, and 2.7 billion characters. Uncompressed, it occupies 2.8 gigabytes of disk space.

In the realm of pretraining corpora, this is a small dataset. Our focus in curation was to produce a corpus free from redundancies (which is not often a focus\footnote{For example, Pile of Law \cite{pileoflaw} contains U.S. state and territory code snapshots, 
for each year for which the data exists on \url{justia.com}. This data comprises a 6.7 GB subset of the 256 GB dataset. Removing verbatim redundancies from this subset, resulting from code that remains unchanged year to year, reduces its size by an order of magnitude.}), and to source text primarily from reputable, authoritative sources. 
An ultimate aim beyond the scope of this work is to produce a minimally sufficient, authoritative corpus for related adjudication tasks. The dataset is intended to support both pretraining and retrieval.

\subsection{Composition}
\label{composition-breakdown}

We describe the composition of our dataset via a partition into six categories. Each document in our corpus belongs to exactly one of the following categories:

\begin{enumerate}
    \item \textbf{Legal}. Current or former U.S. law.
        
    \item \textbf{Regulatory Guidance}. Guidance on U.S. law, released by agencies.

    \item \textbf{Coverage Rules, Contracts, and Medical Policies}. Text outside formal law that describes 
    binding coverage rules\footnote{In some cases law references but does not contain such rules 
    (e.g. state Medicaid handbooks).}. This includes text from contracts, and contract-referenced or proprietary medical policies.
        
    \item \textbf{Opinion, Policy, and Summary}. Opinions, policy perspectives, or summaries of law, 
    proposed law, executive actions, or compliance.
        
    \item \textbf{Case Descriptions}. Reviews of individual health insurance coverage decisions.
        
    \item \textbf{Medical Guidelines and Literature}. Clinical guidelines and medical literature, 
    excluding contract-specific medical policy falling into category 3.
        
\end{enumerate}

For the interested reader we provide a more detailed description of the constituents of the dataset in Appendix \ref{sec:detailedcomposition}.

\subsection{Document-Level Tags}
We equip each document in our corpus with a set of plain text tags. For example, documents comprised of text from U.S. law carry a ``legal" tag. These tags support use as a knowledge base. For example the tags generally support constrained retrieval. This applies in particular to Retrieval Augmented Generation (RAG) \cite{rag}, where one can use the tags to limit the collection of document chunks considered for injection in a prompt. Exact tag-based filtering supports pipelines capable of providing certain types of performance guarantee.

\subsubsection{Partitioning Tags}

We formulated a privileged set of \emph{partitioning tags}, i.e. tags with the property that each document in the dataset belongs to exactly one tag. The privileged set corresponds to the breakdown described in Section \ref{composition-breakdown}. The associated plain text tags are: ``legal'', ``regulatory-guidance'', ``contract-coverage-rule-medical-policy'', ``opinion-policy-summary'', ``case-description'', and ``clinical-guidelines''. 

\subsubsection{Knowledge Base Tag}

We also make use of a knowledge base tag (``kb'') to indicate that a document is authoritative. This is of course an ambiguous determination. Nonetheless, our goal was to formalize this association as a step toward supporting applications requiring authoritative retrieval.

Knowledge base tags were auto generated during dataset curation with a human in the loop. We tried to assign the tag only to documents which are binding in law, or which constitute legal guidance from government agencies. However, even this characterization is ambiguous. For example, a contract is binding in law for those who are parties to the contract, but not for others. Documents which are binding in law in some but not all contexts, such as contracts, were decorated with the `kb' tag.

For example, we label all Medicare Coverage Determination documents with the knowledge base tag. This is because law stipulates the authority of these determinations. As a result they hold established and verifiable weight in individual case adjudications. 

As a non-example, we omit the knowledge base tag for congressional testimony documents. While such testimony is often reputable, and accurate, it is neither binding nor is it formal administrative guidance.

\subsubsection{Example Usage}

Consider a caseworker seeking information on coverage rules for a Medicaid beneficiary in New York state. Here a caseworker could
mean a pharmacist, nurse, physician, attorney, patient advocate, or other individual who is supporting the beneficiary in understanding
and accessing their coverage.
A pipeline can employ filtering to restrict attention to documents with ``legal'', ``kb'', ``new-york'' and ``medicaid'' tags. This provides a simple and efficient mechanism to deterministically avoid retrieval of irrelevant documents. For example, those corresponding to Ohio state law.

This capability is important for at least two reasons. First, it removes the necessity to filter documents using only their semantic content. This allows for more efficient retrieval, which makes deployment 
more accessible for low-resource efforts, and improves latency. Second, tag based filtering supports provable guarantees about the relevance of retrieved documents. While the end to end utility of such guarantees in typical \emph{generative} pipelines is tenuous, the utility in extractive or otherwise guardrailed pipelines need not be.

\subsubsection{Tag Distribution}

\begin{figure}[ht]
\centering
\includegraphics[width = \columnwidth]{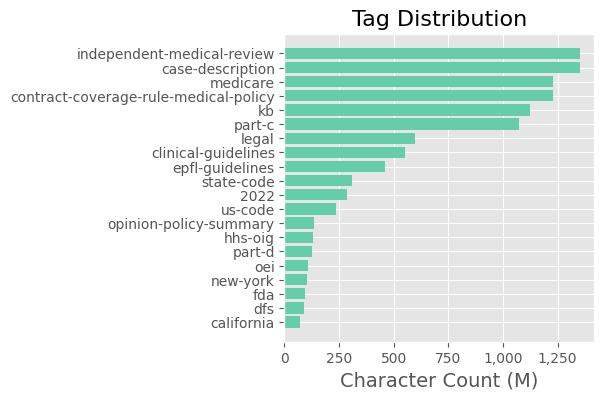}
\caption{Character counts for the twenty most common tags in our dataset. There is a long tail of less common tags.}
\label{tag-distribution-char}
\end{figure}

Table \ref{overall-stats} shows high level statistics about the tag distribution. In total there are 616 unique tags used across the 8,311 documents. Figure \ref{tag-distribution-char} shows the distribution of tags by associated character count.

    \begin{table*}[ht]
\centering
\begin{tabular}{lllll}
\hline
\textbf{Category} & \textbf{Documents} & \textbf{Words (M)} & \textbf{Chars (M)} & \textbf{Size (GB)} \\
\hline
legal & 330 & 86 & 545 & 0.57 \\
\hline
regulatory-guidance & 1,110 & 5 & 38 & 0.04 \\
\hline
contract-coverage-rule-medical-policy & 7 & 196 & 1,228 & 1.31 \\
\hline
opinion-policy-summary & 2,093 & 19 & 131 & 0.14 \\
\hline
case-description & 2,630 & 214 & 1,351 & 1.44 \\
    \hline
clinical-guidelines & 2,150 & 82 & 554 & 0.57 \\
\hline
[kb] & [1,434] & [171] & [1,123] & [1.16] \\
\hline

\textbf{Total} & \textbf{8,311} & \textbf{419} & \textbf{2,703} & \textbf{2.83} \\
\end{tabular}
\caption{\label{overall-stats} Dataset statistics broken down by our privileged partitioning set. We also show statistics for the subset of documents tagged with the knowledge base tag (``kb"). The \textbf{Total} row indicates the sum of the rows excluding the [kb] row.}
\end{table*}

\section{An Appeal Adjudication task}

We now introduce an appeal adjudication task, and an associated dataset.

\subsection{Task Definition}

The task is to predict whether an external appeal of a health insurance coverage denial will result in a full or partial overturn, given a description of the denial context and
two optional pieces of metadata. 
We formulate this as a three class classification problem, with classes corresponding to: full or partial overturns, upheld decisions, and insufficient descriptions. We do not 
require that the description be complete, consist of clinical notes, or be written by experts. The optional metadata specify insurance
 type and regulatory jurisdiction classifications\footnote{In a given implementation, insurance types could take the form of high-level categorizations (e.g. ``Medicaid"), or be more fine grained (e.g. ``New York Medicaid Advantage"). Similarly one could use a high-level categorization of the primary regulatory jurisdiction, such as ``New York", or ``Employee Benefits Security Administration", or a more nuanced and accurate multi-label scheme.} - two factors that influence coverage rules and appeal outcomes.

We define an external appeal to be one submitted to an independent third party. This is in contrast to an internal appeal, which is one submitted to the insurance administrator that issues a denial. External appeal processes exist for most types of health insurance, but vary in nature across them.

\subsection{Motivation}

This task is important for patients and caseworkers engaged in appeals processes. The likelihood of appeal success informs strategy, expectations, and cost-benefit analyses for appellants. It can also help streamline costly administrative work, and inform oversight.

It is important for regulators, who typically want to ensure that external appeal overturn rates are low\footnote{A Medicare star rating measure reflects this goal. See e.g. measure C29 of the \href{https://www.cms.gov/files/document/2024-star-ratings-technical-notes.pdf}{2024 technical notes}.}. Low overturn rates reflect consistency between insurer and third party adjudication, which is some form of fairness.

We compiled a labeled dataset for this task by exploiting historical case outcomes. Those outcomes are present in the ``case-description'' subset of our corpus.

\subsection{Prediction}

Although our labeled benchmark dataset is generated from historical case outcomes, we note that we are singularly focused on supporting actual forecasting. 
That is, we aim to support predicting appeal adjudication outcomes 
in real world situations in which the outcomes have yet to occur. This means our task must be addressed using only information which would be known 
by appellants or their advocates at that time. While this goal may sound obvious or implied from our setup, there has 
recently been  critical and justified examination of misuse of the term `prediction' in legal language tasks that
perform classification using information that is not realistically available in the real world settings they purportedly support \cite{legal_judgement_prediction_do_it_right,rethinking_automatic_prediction}.

In the following sections we introduce a process for learning to extract subsets of case summaries to obtain context
which is reflective of that known by patients prior to their appeals. In this way, our approach to the task is truly one of
of \emph{outcome forecasting}, rather than post hoc \emph{outcome-based judgement identification}, to use the language of \cite{rethinking_automatic_prediction}.
This is because our process is designed to extract summaries that only contain statements which could be known before an appeal is submitted.

Our focus on this quality of our inputs is central to our formulation, as we are interested in real
world use of our models to support patients in accessing coverage to which they are entitled, rather than retrospective classification models tailored to
the qualities of review summaries.

\subsection{Challenges}

Our task is difficult for many reasons. 

Most notable is that outcomes may depend on facts not present in our model inputs, such as jurisdiction, governing law, contracts, medical history, policy, and literature. These vary by state, insured, insurer, and insurance type, and change over time.

This renders our modeling task under specified in many contexts, with no correct answer for many inputs. For example, what should a model predict when an appeal would be overturned under Medicare rules but upheld under New York Medicaid rules, if insurance type isn't specified?

We introduce our task despite this difficulty for three reasons. 

First, responsible deployment of a well-performing model would be valuable and pose few risks. Although such a model cannot be perfect, it can help improve access to justice. For some contexts, denials are overwhelmingly inappropriate, across jurisdictions, insurance types, and contracts. It is possible to identify such context as \emph{likely} to be problematic. This is useful in the same way a prediction for 90\% chance of rain (another theoretically under specified forecasting task) is useful. It would be irresponsible to bet one's wellbeing on the prediction, but carrying an umbrella is a relatively harmless precaution. Models that excel at our task can help promote access to justice, without posing risks, if deployed with caution and with limited scope.

Second, we hope this work promotes further research on related tasks. For example, one could extend our benchmark to additionally include optional input metadata encoding medical billing codes.

Finally, our task circumvents a common difficulty for patients: accessing important details. Patients face significant barriers in accessing important details. They may not have full access to contracts or medical records, and may not know who funds their plan. These details are critically relevant in helping patients navigate recourse available to them. Nonetheless, they are often inaccessible, even when law ostensibly protects access. Accurate prediction tools with lower barriers to entry can improve access to justice for those who might otherwise forgo recourse.

It is important to note that our task differs from that performed by adjudicators. Reviewers rely on charts, notes, diagnoses, and rationales, and can request more information. Our benchmark descriptions provide only brief explanations which are always insufficient for adjudication, and sometimes also insufficient for high quality outcome forecasting.

Some explanations in our benchmark are too general to hold any predictive power. For example, a 
summary might say ``My hospital stay was denied''. There are many reasons for hospital admission, 
and hospital admission is routinely rightfully \emph{and} wrongfully denied.

At the other extreme, some summaries can be used to predict appeal adjudication outcomes effectively. This can be because 
they contain specific detail, or describe situations that typically yield consistent outcomes regardless of details. 
For example, among state regulated commercial plans in California, more than 80\% of appealed denials for Harvoni 
for Hepatitis C treatment are overturned \cite{persiusCaliforniaExternalAppeal}.

This variability is both challenging and reflects a real problem encountered in manual casework. Patients 
request help with varying specificity. Modeling whether a request is sufficient 
to predict an outcome is important for applications. We want to avoid making predictions 
when there is insufficient information, which is why the problem is cast as one of three class classification.

We discuss more subtleties of our particular task formulation in Appendix \ref{sec:task-description}.

\subsection{Raw Source Data}

To construct annotated data, we used three sources whose unlabeled text are also present in our corpus.

These sources are:

\begin{enumerate}
    \item A New York DFS External Appeal Database.
    \item A California CDI External Appeal Database.
    \item A California DMHC External Appeal Database.
\end{enumerate}

Appendix \ref{sec:detailedcomposition} has more information about these sources.

The sources contain plain text descriptions of external appeal cases, often including background, summaries of the outcomes, and rationales for decisions. Separately, structured metadata is reported.
We extracted from the structured metadata a rudimentary insurance type and jurisdiction classification scheme: each case is labeled with an insurance type among 
\verb|{Commercial, Medicaid, Unspecified}|, and a jurisdiction among \verb|{NY, CA, Unspecified}|\footnote{Because the insurance types
and jurisdictions do not vary much across our data, we do not expect them to be impactful for modeling \emph{in our benchmark}. We do expect them to be useful for the general task.}.

The standards used to structure the plain text descriptions are inconsistent both between and within the sources. Typically, the adjudication rationale leaks information about the outcome, whereas the background does not. 

Together, all these features make it nontrivial to extract useful (description, outcome) pairs, as meaningful training and evaluation  of forecasting models requires descriptions which do not leak the outcome.

Figure \ref{argilla-annotation-task} illustrates the issues. We reiterate that across our data sources, case descriptions leak case outcomes in varied ways. This example illustrates one of many case description formats and mechanisms of leaking. The high degree of variability makes immediate solutions like splitting on keywords ineffective.

\begin{figure*}[ht]

\begin{tcolorbox}[colback=blue!3!white, colframe=blue!35!black, title=Source Text]
\textcolor{BrickRed}{The parents of a seventeen month-old male with respiratory distress requiring intubation, surfactant, and episodes of supraventricular tachycardia (SVT) are requesting reimbursement and prospective authorization for Synagis (pallivizumab).} Reviewers Findings: The reviewing physician found that RSV infection has been associated with tachyarrhythmias including SVT that may be hemodynamically significant in infants such as this one. The reviewing physician found that the Health Plan denial should be overturned.
\end{tcolorbox}

\caption{ An example of a raw case description from our unlabeled corpus. Background context that summarizes the facts of the case knowable prior to its review by an external reviewer is indicated in red. The subsequent non-background text leaks the case outcome determination. We used \href{https://github.com/argilla-io/argilla}{argilla} to manually annotate background context. }

\label{argilla-annotation-task}
\end{figure*}

\subsection{Construction of Training Data}
\label{training-data-construction}
To construct a suitable training dataset, we first produced manual span annotations for a subset of case descriptions. We labeled subspans encoding background context one would typically know before an appeal submission, and scored the spans on a scale from 1 to 4, based on the degree to which they would be sufficient to make an informed prediction about the adjudication outcome in isolation.

We then used this annotated subset to train two `bootstrapping' models. 
One is a span selection model responsible for extracting background from potentially leaking case descriptions. 
The other is a binary classifier for such background text, indicating whether the text is sufficient in isolation to predict a case outcome. 
We used the models together to perform extraction of non-leaking background from all 73,987 of our case descriptions, and to endow them with three-class pseudolabels. 
Finally, we used this larger collection of pseudo-annotated case summaries to train appeal outcome classifiers.

\subsubsection{Manual Annotation}

We manually annotated spans from 1,000 case descriptions, with one third coming from each of the three data sources. The descriptions from each data source were sampled randomly. 
We used argilla \cite{argilla} to annotate non-leaking background spans and label those spans with sufficiency scores. 

We intend to improve the corpus and benchmark detailed in this work, and we are currently working to streamline and iteratively improve our annotation guidelines, and to 
generate more manual annotations. Encouraged by our initial modeling results, described in more detail in what follows, we plan to enlist more professionals 
experienced in casework to each manually annotate the same set of 3,000 case descriptions. We will then measure inter-annotator agreement for spans
and sufficiency scores via numerous measures, to evaluate the extent to which our manual annotation process yields reliable, and consistent results.
At present, we have no such consistency results to share.

More details about the guidelines provided to annotators are shared in Appendix \ref{sec:annotation-guidelines}.

\subsubsection{Span Selector Training}

We trained a span selection model to extract non-leaking background context from our case descriptions. 
We finetuned Distilbert \cite{distilbert} for this task, and found subjectively\footnote{We evaluated our span selector quantitatively, but our determination that the model was adequate 
was a subjective one based on qualitative human evaluation of many samples. The metrics for token-level classification on a held out test set
were 89.8\% recall, 92.2\% precision, 91.0\% F1, and 93.2\% accuracy.} that 
it gave results of sufficient quality. We did not experiment with alternative approaches to span selection 
once we successfully trained the Distilbert variant\footnote{Our first approach was to use a language model to perform the requisite span selection. We found initial results with gpt-4o and Claude 3 variants unsatisfactory. Given the success of this cheap approach which also affords perpetual, negligible-cost access to the model artifacts, we did not pursue that path further.}.

\subsubsection{Sufficiency Model Training}

We trained a model to classify background as sufficient or insufficient to predict a case outcome. We also finetuned Distilbert for this task. 
We produced binary labels from our ranked sufficiency score annotations by labeling all scores less than 
3 as ``Insufficient'', and all scores greater than or equal to 3 as ``Sufficient''. This choice was informed by our current annotation guidelines,
which are documented in Appendix \ref{sec:annotation-guidelines}. In short, scores less than 3 reflect descriptions which
do not specify either the service, or the diagnosis or medical issue for which the service is being provided. Without both of these pieces of
information, it is impossible to assess the likelihood of appeal overturn with any fidelity.

\subsubsection{Bootstrap Model Application}

To complete construction of the dataset, we applied our trained span selection model to 73,987 case descriptions, and 
then applied our sufficiency classifier. We partitioned the results into training and test splits; splits were stratified by 
source and appeal outcome and otherwise determined via uniform sampling.

An example of a non-leaking case summary extracted by our trained model artifact is displayed in Figure \ref{case-summary-figure}.

\begin{figure}[ht]

    \begin{tcolorbox}[colback=green!3!white, colframe=green!35!black, title=Extracted Case Summary]
    A 52-year-old female enrollee has requested reimbursement for the Decision Dx melanoma assay provided on 11/18/15. The Health Insurer has denied this request indicating that the testing at issue was considered investigational for evaluation of the enrollee's malignant melanoma.
        \end{tcolorbox}
    \caption{An example of non-leaking background context extracted from a full, leaking case description, by our trained Distilbert-based span selector.}
    	\label{case-summary-figure}
\end{figure}

\subsection{Baselines}

We trained end to end outcome prediction models on our benchmark to serve as baselines for further modeling efforts. The choice of models we evaluated was, as is often the case, essentially arbitrary.

Table \ref{model-performance} shows the results of evaluating a set of finetuned models on the test set of our benchmark. 

The models tested include pretrained bidirectional transformers (BERT and DistilBERT variants).  
One model, ClinicalBERT \cite{clinicalbert}, was already pretrained on medical corpora. Another, 
LegalBERT \cite{legalbert}, was already pretrained on legal corpora.

We also compared performance to a popular closed source language model, GPT-4o-mini.
 GPT-4o-mini was provided a system prompt explaining the task, and a required output format. 
 It was then evaluated in a two shot context. The best performing model among those 
 we tested was the Distilbert variant. Table \ref{model-performance} records the results on a per-metric basis.

\begin{table*}[h]
\centering
\begin{tabular}{p{11em}llllll}
\hline
\textbf{Model} & \textbf{Accuracy} & \textbf{Precision} & \textbf{Recall} & \textbf{F1}  & \textbf{ROC AUC} & \textbf{Params (M)} \\
\hline
gpt-4o-mini-2024-07-18 (2-shot) & .532 & .499 & .565 & .397 & N/A & unknown \\
    \hline
    legalbert-small-uncased & .707 & .753 & .669 & .704 & .857& 35\\
\hline
distilbert-base-uncased & .733  & \textbf{.772}  & \textbf{.695}  & \textbf{.727} & .878 & 67\\
\hline
clinical-bert & \textbf{.740} & .758 & .688 & .717 & \textbf{.889}  & 135  \\
\end{tabular}
\caption{Classification metrics on the held out test set of our case adjudication benchmark. Each metric is represented by a macro average across the three classes, 
and the best performing model for each macro metric is in bold. Note that the macro F1 is not the harmonic mean of the macro recall and macro precision. In the same vein, it is not true that the value in the F1 column must fall between
the values presented in the Precision and Recall columns, as seen in the gpt metrics. The test set is comprised of 9,745 records, of which 175 are labeled ``Insufficient", 5,177 are labeled ``Upheld", and 4,393 are labeled ``Overturned".}
\label{model-performance}
\end{table*}

Our training and evaluation methodologies are detailed further in Appendix \ref{sec:experimental-setup}.

\section{Discussion}
\label{discussion}
 We now turn to discussing potential applications of this work.

\subsection{Intended Use}

Our corpus, task, and benchmarks were all constructed with downstream applications in mind. We list a few.

\textbf{Corpus as a Partial Knowledge Base}
Health insurance coverage rules, contracts, and medical policies are often complex. They are difficult to efficiently and accurately apply to the facts of particular situations, even for experts. Existing generative tools are capable of summarizing the complexities of individual provisions, but to do so they need access to authoritative ground truth. Many deployed solutions suffer from both insufficient access to such ground truth, and an abundance of access to non-authoritative or inaccurate text.

Our introduction of this corpus is a step towards a minimally sufficient knowledge base. 
It can support generative and extractive tasks related to understanding of U.S. health insurance. It is particularly
well suited to legal question answering for health insurance related questions.

\textbf{Models as Oversight Tools}
Regulators could use high quality appeal outcome predictors to promote improved outcomes and adjudication. 
For example, they could require that cases with high likelihood of overturn get reviewed in shortened time frames. 
This could improve mean time to access-to-care for overturned pre-service denials, which has a major impact on health outcomes \cite{kffDenialSurvey,commonwealthUnforeseenCostsArticle}.

\textbf{Models as Patient Self Help Tools}
Appeal outcome predictors can support patients who do not have access to expert support.

Blind reliance and automated use carries risk, and we do not recommend such use. 
Nonetheless, these models can play effective and safe supporting roles.

Blind reliance has problematic implications. Models could incorrectly convince a patient they can't win an appeal. If this led to a patient unnecessarily forgoing care, the consequences could be dire. This would be especially problematic when predictions differ from those that would be provided by human experts.

Risk mitigation is possible through worst case scenario assessment. 
By designing self-help applications for responsible use, acceptable risk levels are achievable.
 One mechanism is designing to promote `qualified trust' - that is, granting trust qualified by explicit risk assessment.

For example, applications could instruct those desperately needing coverage to ignore predictions and seek the support of human experts. 
Many free and low-cost support systems exist across the U.S. In such cases, prediction inaccuracy risks outweigh benefits.

When lack of coverage has important, but not dire, implications, applications could instruct patients to seek expert advice. For those without access to such help, the models can serve as fallbacks. The risk of prediction inaccuracy might or might not outweigh the risks of having no support at all. 
For example, pursuing an appeal that is ultimately unsuccessful costs time. Forgoing a post-service appeal and paying for care in a case that would have resulted in overturn costs money. 
Forgoing a pre-service appeal and forgoing care as a result in a case that would have resulted in overturn poses risks to one's health. The risk tradeoff is subjective, and case dependent.

When lack of coverage has minimal implications, our models can safely support patients most easily.
Patients in such scenarios often 
forgo post-service appeals due to the mistaken belief that they are unlikely to win. 
The models provide a quick way for patients to update their belief in the likelihood of appeal success. 

Self-help tools designed through a lens of qualified trust can improve patient outcomes.

\textbf{Task as an Application Entry Point}
Our task has notable shortcomings and limitations. However, the more general conceptual context merits further exploration. 

Our task illuminates the need for progress on a suite of related tasks. We hope that our task serves as an entry point for other researchers to contribute to this domain. Our adjudication data and annotation workflow provide a straightforward path to such contributions.

We list some alternate and complementary tasks of interest.

\begin{enumerate}
    \item \textbf{It would be useful to predict whether an appeal \emph{merits} overturn, according to law or policy.}
    \label{alt-task} Such a model could help regulators and administrators triage appeal case loads. Effective triage could lower mean-time-to-case-resolution for \emph{inappropriate} denials. A model of this type could also help regulators audit review outcomes. One could approach this task with our methods and alternate annotation guidelines.
    
    \item \textbf{It would be useful to cite evidence supporting predictions for the task in item \ref{alt-task}.} 
    A simple formulation could use a language model and RAG pipeline to achieve this. It is straightforward to use our corpus as a partial knowledge base towards this end.

    \item \textbf{It would be useful to construct complete, coherent arguments supporting predictions for the task in item \ref{alt-task}.}
    This could take the form of generating `proof' for an adjudication determination. Such proof could be primarily extractive in nature, relying on authoritative references. It would require a dynamic determination of a minimally sufficient set of source chunks, unlike a typical RAG pipeline, and coherent assimilation of those sources. The exact approach could take many forms (see e.g. \cite{proofwriter} for one motivating example).

    \item \textbf{It would be useful to draft high quality appeal letters for patients and providers.}
    Appeal processes are laborious, and present many barriers. One intensive step is writing a 
    letter arguing for overturn. Some reviewers consider letters minimally, rendering generic or inaccurate letters harmless. However, letter quality and accuracy 
    become critical in dire cases or when reviewers might miss important context. Deploying low quality letter generators without precautions harms the most vulnerable.
    Nonetheless, it is trivial to take this approach, and many companies are doing so with varied (monetary) success. 
    
    Responsibly deploying letter generators that present accurate and verifiable evidence is more difficult. 
    Models with this capability offer many benefits. This task is of course deeply related to the previous task.
\end{enumerate}

\textbf{Benchmark as a Minimal Prototype}
Our appeal adjudication task formulation reflects a realistic application with broad potential impact. 
On the other hand, our preliminary \emph{benchmark dataset} is a rudimentary formulation of the task. 
The corpus and task definition, rather than the task benchmark and model implementations, 
are the more central concerns of this work. The task data and models serve as minimal, first iteration prototypes 
for a more general research need. We hope the benchmarks will inspire further research and more refined formulations 
of benchmark data of the same flavor in this space.

\subsection{Out of Scope Use}

There are many ways to use this work that are not intended or advised. These include:

\begin{itemize}

\item Treating the unlabeled corpus as a \emph{complete} knowledge base.
\item Applying our benchmark models to alternate tasks (e.g. interpreting the outputs as appropriate, rather than expected, outcomes).
\item Using our data to fully automate claims denial review and recourse processes.

\end{itemize}

Some out of scope uses of our work arise from inherent limitations, which we discuss in more detail in Section \ref{limitations}. Others arise from associated risks. For example, it is possible to use this work to deploy applications that would negatively affect patient outcomes, despite our intent. We discuss the risk landscape in Section \ref{ethics}.

\section{Conclusion}

In this work we introduced a corpus of reputable text pertaining to health insurance coverage rules, and discussed potential applications. We also introduced a case outcome prediction task, and a benchmark dataset for that task. We trained and evaluated model baselines on our introduced benchmark. The data and task show promise for improving efficient understanding of health insurance coverage rules.

\section{Limitations}
\label{limitations}

\subsection{Task Shortcomings}
Our appeal adjudication task has room for improvement; for example, it could support additional optional metadata. Appeal outcomes often depend on medical records, billing codes, and other missing factors.

\subsection{Simplicity of the Benchmark}

As detailed in Section \ref{training-data-construction}, we bootstrapped our benchmark from 1,000 manual annotations. Furthermore, we extracted background descriptions from a set of relatively short case descriptions. It would be beneficial to draw descriptions from a more diverse distribution, and to perform more manual annotation.

\subsection{Corpus Deficiencies}
Our corpus lacks information necessary to make certain types of coverage determinations. This includes commercial insurance contracts,
medical policies, Medicaid and CHIP state plans, Medicare managed care contracts, and the most recent state and U.S. code.

\section{Ethical Considerations}
\label{ethics}

\subsection{Potential for Propagation of Bias}

The risk of propagating harmful bias is central in machine learning. One hopes to learn about a target problem from training data, but risks instead learning about bias in that data. This is particularly problematic when tools are widely used before they are evaluated for such bias.

The proliferation of language models has worsened these problems \cite{llm-bias-propagation}. It is easier than ever to develop, deploy, and gain adoption for tools with minimal effort. At the same time, language models have less built-in guarantees than many other ML models. Their evaluation also requires more subtle, subjective choices.

Healthcare applications are particularly fraught. There are already stark disparities and vast collections of data reflecting those disparities. Automated tools have consistently led to bias propagation \cite{dissecting_racial_bias, llms-propagate-race-based-medicine}.

There are inequities in U.S. healthcare associated with race, sex, gender identity, age, disease states, disability status, and income, among many other things. Our work could propagate such biases through unintended use.

Particular concerns include:

\begin{enumerate}
\item Conflating historical trends with \emph{just} or \emph{appropriate} trends.

This concern informed our task formulation. Modeling whether a case \emph{merits} overturn with our data would be misguided because historical outcome data encodes biases we cannot sufficiently account for.

\item Demographic selection bias and inequitable benefit.

Our case data has minimal demographic information, and mostly corresponds to individuals over the age of 50 with commercial plans in New York and California. It also reflects only externally appealed denials, rather than the complete distribution. Information about racial, ethnic, and income distributions is limited, and more robust evaluations are necessary. Models trained on our benchmark will perform best in contexts common in the data, potentially exacerbating disparities.
    
\end{enumerate}

\subsection{Potential for Misuse}

There is also potential for our data or models to worsen patient outcomes in ways unrelated to bias. This could occur through reckless use, or intentionally (e.g. for financial gain). 

Organizations quickly deploy prototypes to raise funding and acquire customers. A reckless entrepreneur could use our work to deploy immature patient support tools. Without guardrails or clear disclaimers about risks, such use could have disastrous consequences. Many people who need support are fighting for coverage of care critical to their well-being.

More concerning is the possibility of nefarious misuse. Corporations with extensive resources are deploying ML tools to lower their costs. When done legally, and responsibly, there are many benefits to society. But healthcare organizations have used ML based cost-cutting measures irresponsibly \cite{stat-series}. Such irresponsible misuse is a major risk. For example, one could repurpose our overturn prediction models as semi-automated appeal reviewers.

\bibliography{references}

\begin{thebibliography}{53}
\expandafter\ifx\csname natexlab\endcsname\relax\def\natexlab#1{#1}\fi

\bibitem[{{Alsentzer} et~al.(2019){Alsentzer}, {Murphy}, {Boag}, {Weng}, {Jin}, {Naumann}, and {McDermott}}]{clinicalbert2}
Emily {Alsentzer}, John~R. {Murphy}, Willie {Boag}, Wei-Hung {Weng}, Di~{Jin}, Tristan {Naumann}, and Matthew B.~A. {McDermott}. 2019.
\newblock \href {https://doi.org/10.48550/arXiv.1904.03323} {{Publicly Available Clinical BERT Embeddings}}.
\newblock \emph{arXiv e-prints}, page arXiv:1904.03323.

\bibitem[{Aracena et~al.(2023)Aracena, Rodr{\'\i}guez, Rocco, and Dunstan}]{spanish_insurance_coverage}
Claudio Aracena, Nicol{\'a}s Rodr{\'\i}guez, Victor Rocco, and Jocelyn Dunstan. 2023.
\newblock \href {https://doi.org/10.18653/v1/2023.clinicalnlp-1.46} {Pre-trained language models in {S}panish for health insurance coverage}.
\newblock In \emph{Proceedings of the 5th Clinical Natural Language Processing Workshop}, pages 433--438, Toronto, Canada. Association for Computational Linguistics.

\bibitem[{{Bauer} et~al.(2023){Bauer}, {Stammbach}, {Gu}, and {Ash}}]{legal-extractive}
Emmanuel {Bauer}, Dominik {Stammbach}, Nianlong {Gu}, and Elliott {Ash}. 2023.
\newblock \href {https://doi.org/10.48550/arXiv.2305.08428} {{Legal Extractive Summarization of U.S. Court Opinions}}.
\newblock \emph{arXiv e-prints}, page arXiv:2305.08428.

\bibitem[{Chalkidis et~al.(2019)Chalkidis, Androutsopoulos, and Aletras}]{neural-legal-judgement-pred}
Ilias Chalkidis, Ion Androutsopoulos, and Nikolaos Aletras. 2019.
\newblock \href {https://doi.org/10.18653/v1/P19-1424} {Neural legal judgment prediction in {E}nglish}.
\newblock In \emph{Proceedings of the 57th Annual Meeting of the Association for Computational Linguistics}, pages 4317--4323, Florence, Italy. Association for Computational Linguistics.

\bibitem[{{Chalkidis} et~al.(2020){Chalkidis}, {Fergadiotis}, {Malakasiotis}, {Aletras}, and {Androutsopoulos}}]{legalbert}
Ilias {Chalkidis}, Manos {Fergadiotis}, Prodromos {Malakasiotis}, Nikolaos {Aletras}, and Ion {Androutsopoulos}. 2020.
\newblock \href {https://doi.org/10.48550/arXiv.2010.02559} {{LEGAL-BERT: The Muppets straight out of Law School}}.
\newblock \emph{arXiv e-prints}, page arXiv:2010.02559.

\bibitem[{{Chalkidis} et~al.(2023){Chalkidis}, {Garneau}, {Goanta}, {Katz}, and {S{\o}gaard}}]{legalllama}
Ilias {Chalkidis}, Nicolas {Garneau}, Catalina {Goanta}, Daniel~Martin {Katz}, and Anders {S{\o}gaard}. 2023.
\newblock \href {https://doi.org/10.48550/arXiv.2305.07507} {{LeXFiles and LegalLAMA: Facilitating English Multinational Legal Language Model Development}}.
\newblock \emph{arXiv e-prints}, page arXiv:2305.07507.

\bibitem[{{Chalkidis} et~al.(2021){Chalkidis}, {Jana}, {Hartung}, {Bommarito}, {Androutsopoulos}, {Katz}, and {Aletras}}]{lexglue}
Ilias {Chalkidis}, Abhik {Jana}, Dirk {Hartung}, Michael {Bommarito}, Ion {Androutsopoulos}, Daniel~Martin {Katz}, and Nikolaos {Aletras}. 2021.
\newblock \href {https://doi.org/10.48550/arXiv.2110.00976} {{LexGLUE: A Benchmark Dataset for Legal Language Understanding in English}}.
\newblock \emph{arXiv e-prints}, page arXiv:2110.00976.

\bibitem[{{Chen} et~al.(2023){Chen}, {Hern{\'a}ndez Cano}, {Romanou}, {Bonnet}, {Matoba}, {Salvi}, {Pagliardini}, {Fan}, {K{\"o}pf}, {Mohtashami}, {Sallinen}, {Sakhaeirad}, {Swamy}, {Krawczuk}, {Bayazit}, {Marmet}, {Montariol}, {Hartley}, {Jaggi}, and {Bosselut}}]{meditron}
Zeming {Chen}, Alejandro {Hern{\'a}ndez Cano}, Angelika {Romanou}, Antoine {Bonnet}, Kyle {Matoba}, Francesco {Salvi}, Matteo {Pagliardini}, Simin {Fan}, Andreas {K{\"o}pf}, Amirkeivan {Mohtashami}, Alexandre {Sallinen}, Alireza {Sakhaeirad}, Vinitra {Swamy}, Igor {Krawczuk}, Deniz {Bayazit}, Axel {Marmet}, Syrielle {Montariol}, Mary-Anne {Hartley}, Martin {Jaggi}, and Antoine {Bosselut}. 2023.
\newblock \href {https://doi.org/10.48550/arXiv.2311.16079} {{MEDITRON-70B: Scaling Medical Pretraining for Large Language Models}}.
\newblock \emph{arXiv e-prints}, page arXiv:2311.16079.

\bibitem[{Collins et~al.(2023)Collins, Roy, and Masitha}]{commonwealthDebtArticle}
Sarah Collins, Shreya Roy, and Relebohile Masitha. 2023.
\newblock \href {https://www.commonwealthfund.org/publications/surveys/2023/oct/paying-for-it-costs-debt-americans-sicker-poorer-2023-affordability-survey} {Paying for it: How health care costs and medical debt are making americans sicker and poorer}.
\newblock Accessed on January 3, 2024.

\bibitem[{Daniel and Francisco(2023)}]{argilla}
Vila-Suero Daniel and Aranda Francisco. 2023.
\newblock \href {https://github.com/argilla-io/argilla} {{Argilla - Open-source framework for data-centric NLP}}.

\bibitem[{Dhamala et~al.(2021)Dhamala, Sun, Kumar, Krishna, Pruksachatkun, Chang, and Gupta}]{llm-bias-propagation}
Jwala Dhamala, Tony Sun, Varun Kumar, Satyapriya Krishna, Yada Pruksachatkun, Kai-Wei Chang, and Rahul Gupta. 2021.
\newblock Bold: Dataset and metrics for measuring biases in open-ended language generation.
\newblock In \emph{Proceedings of the 2021 ACM conference on fairness, accountability, and transparency}, pages 862--872.

\bibitem[{{Elwany} et~al.(2019){Elwany}, {Moore}, and {Oberoi}}]{bertlawschool}
Emad {Elwany}, Dave {Moore}, and Gaurav {Oberoi}. 2019.
\newblock \href {https://doi.org/10.48550/arXiv.1911.00473} {{BERT Goes to Law School: Quantifying the Competitive Advantage of Access to Large Legal Corpora in Contract Understanding}}.
\newblock \emph{arXiv e-prints}, page arXiv:1911.00473.

\bibitem[{Gartner(2023)}]{persiusCaliforniaExternalAppeal}
Michael Gartner. 2023.
\newblock \href {https://blog.persius.org/blog/ca-external-appeals-demographics} {California external appeal outcome demographics}.
\newblock Accessed on November 15, 2024.

\bibitem[{{Gu} et~al.(2020){Gu}, {Tinn}, {Cheng}, {Lucas}, {Usuyama}, {Liu}, {Naumann}, {Gao}, and {Poon}}]{domain-specific-biomedical}
Yu~{Gu}, Robert {Tinn}, Hao {Cheng}, Michael {Lucas}, Naoto {Usuyama}, Xiaodong {Liu}, Tristan {Naumann}, Jianfeng {Gao}, and Hoifung {Poon}. 2020.
\newblock \href {https://doi.org/10.48550/arXiv.2007.15779} {{Domain-Specific Language Model Pretraining for Biomedical Natural Language Processing}}.
\newblock \emph{arXiv e-prints}, page arXiv:2007.15779.

\bibitem[{{Guha} et~al.(2023){Guha}, {Nyarko}, {Ho}, {R{\'e}}, {Chilton}, {Narayana}, {Chohlas-Wood}, {Peters}, {Waldon}, {Rockmore}, {Zambrano}, {Talisman}, {Hoque}, {Surani}, {Fagan}, {Sarfaty}, {Dickinson}, {Porat}, {Hegland}, {Wu}, {Nudell}, {Niklaus}, {Nay}, {Choi}, {Tobia}, {Hagan}, {Ma}, {Livermore}, {Rasumov-Rahe}, {Holzenberger}, {Kolt}, {Henderson}, {Rehaag}, {Goel}, {Gao}, {Williams}, {Gandhi}, {Zur}, {Iyer}, and {Li}}]{legalbench}
Neel {Guha}, Julian {Nyarko}, Daniel~E. {Ho}, Christopher {R{\'e}}, Adam {Chilton}, Aditya {Narayana}, Alex {Chohlas-Wood}, Austin {Peters}, Brandon {Waldon}, Daniel~N. {Rockmore}, Diego {Zambrano}, Dmitry {Talisman}, Enam {Hoque}, Faiz {Surani}, Frank {Fagan}, Galit {Sarfaty}, Gregory~M. {Dickinson}, Haggai {Porat}, Jason {Hegland}, Jessica {Wu}, Joe {Nudell}, Joel {Niklaus}, John {Nay}, Jonathan~H. {Choi}, Kevin {Tobia}, Margaret {Hagan}, Megan {Ma}, Michael {Livermore}, Nikon {Rasumov-Rahe}, Nils {Holzenberger}, Noam {Kolt}, Peter {Henderson}, Sean {Rehaag}, Sharad {Goel}, Shang {Gao}, Spencer {Williams}, Sunny {Gandhi}, Tom {Zur}, Varun {Iyer}, and Zehua {Li}. 2023.
\newblock \href {https://doi.org/10.48550/arXiv.2308.11462} {{LegalBench: A Collaboratively Built Benchmark for Measuring Legal Reasoning in Large Language Models}}.
\newblock \emph{arXiv e-prints}, page arXiv:2308.11462.

\bibitem[{Gupta et~al.(2024)Gupta, Collins, Roy, and Masitha}]{commonwealthUnforeseenCostsArticle}
Avni Gupta, Sara~R. Collins, Shreya Roy, and Relebohile Masitha. 2024.
\newblock \href {https://www.commonwealthfund.org/publications/issue-briefs/2024/aug/unforeseen-health-care-bills-coverage-denials-by-insurers} {Unforeseen health care bills and coverage denials by health insurers in the u.s.}
\newblock Accessed on April 27, 2025.

\bibitem[{Gururangan et~al.(2020)Gururangan, Marasovi{\'c}, Swayamdipta, Lo, Beltagy, Downey, and Smith}]{dont-stop-pretraining}
Suchin Gururangan, Ana Marasovi{\'c}, Swabha Swayamdipta, Kyle Lo, Iz~Beltagy, Doug Downey, and Noah~A. Smith. 2020.
\newblock \href {https://doi.org/10.18653/v1/2020.acl-main.740} {Don{'}t stop pretraining: Adapt language models to domains and tasks}.
\newblock In \emph{Proceedings of the 58th Annual Meeting of the Association for Computational Linguistics}, pages 8342--8360, Online. Association for Computational Linguistics.

\bibitem[{{Habernal} et~al.(2022){Habernal}, {Faber}, {Recchia}, {Bretthauer}, {Gurevych}, {Spiecker genannt D{\"o}hmann}, and {Burchard}}]{mining-legal-arguments-in-court}
Ivan {Habernal}, Daniel {Faber}, Nicola {Recchia}, Sebastian {Bretthauer}, Iryna {Gurevych}, Indra {Spiecker genannt D{\"o}hmann}, and Christoph {Burchard}. 2022.
\newblock \href {https://doi.org/10.48550/arXiv.2208.06178} {{Mining Legal Arguments in Court Decisions}}.
\newblock \emph{arXiv e-prints}, page arXiv:2208.06178.

\bibitem[{{Henderson} et~al.(2022){Henderson}, {Krass}, {Zheng}, {Guha}, {Manning}, {Jurafsky}, and {Ho}}]{pileoflaw}
Peter {Henderson}, Mark~S. {Krass}, Lucia {Zheng}, Neel {Guha}, Christopher~D. {Manning}, Dan {Jurafsky}, and Daniel~E. {Ho}. 2022.
\newblock \href {https://doi.org/10.48550/arXiv.2207.00220} {{Pile of Law: Learning Responsible Data Filtering from the Law and a 256GB Open-Source Legal Dataset}}.
\newblock \emph{arXiv e-prints}, page arXiv:2207.00220.

\bibitem[{{Hendrycks} et~al.(2021){Hendrycks}, {Burns}, {Chen}, and {Ball}}]{cuad}
Dan {Hendrycks}, Collin {Burns}, Anya {Chen}, and Spencer {Ball}. 2021.
\newblock \href {https://doi.org/10.48550/arXiv.2103.06268} {{CUAD: An Expert-Annotated NLP Dataset for Legal Contract Review}}.
\newblock \emph{arXiv e-prints}, page arXiv:2103.06268.

\bibitem[{{Hua} et~al.(2022){Hua}, {Zhang}, {Chen}, {Li}, and {Weber}}]{legalreelectra}
Wenyue {Hua}, Yuchen {Zhang}, Zhe {Chen}, Josie {Li}, and Melanie {Weber}. 2022.
\newblock \href {https://doi.org/10.48550/arXiv.2212.08204} {{LegalRelectra: Mixed-domain Language Modeling for Long-range Legal Text Comprehension}}.
\newblock \emph{arXiv e-prints}, page arXiv:2212.08204.

\bibitem[{{Huang} et~al.(2019){Huang}, {Altosaar}, and {Ranganath}}]{clinicalbert}
Kexin {Huang}, Jaan {Altosaar}, and Rajesh {Ranganath}. 2019.
\newblock \href {https://doi.org/10.48550/arXiv.1904.05342} {{ClinicalBERT: Modeling Clinical Notes and Predicting Hospital Readmission}}.
\newblock \emph{arXiv e-prints}, page arXiv:1904.05342.

\bibitem[{Jiang et~al.(2023)Jiang, Liu, Nejatian, Nasir-Moin, Wang, Abidin, Eaton, Riina, Laufer, Punjabi et~al.}]{health_system_scale_model}
Lavender~Yao Jiang, Xujin~Chris Liu, Nima~Pour Nejatian, Mustafa Nasir-Moin, Duo Wang, Anas Abidin, Kevin Eaton, Howard~Antony Riina, Ilya Laufer, Paawan Punjabi, et~al. 2023.
\newblock Health system-scale language models are all-purpose prediction engines.
\newblock \emph{Nature}, 619(7969):357--362.

\bibitem[{{Jin} et~al.(2020){Jin}, {Pan}, {Oufattole}, {Weng}, {Fang}, and {Szolovits}}]{medQA}
Di~{Jin}, Eileen {Pan}, Nassim {Oufattole}, Wei-Hung {Weng}, Hanyi {Fang}, and Peter {Szolovits}. 2020.
\newblock \href {https://doi.org/10.48550/arXiv.2009.13081} {{What Disease does this Patient Have? A Large-scale Open Domain Question Answering Dataset from Medical Exams}}.
\newblock \emph{arXiv e-prints}, page arXiv:2009.13081.

\bibitem[{{Jin} et~al.(2019){Jin}, {Dhingra}, {Liu}, {Cohen}, and {Lu}}]{pubmedQA}
Qiao {Jin}, Bhuwan {Dhingra}, Zhengping {Liu}, William~W. {Cohen}, and Xinghua {Lu}. 2019.
\newblock \href {https://doi.org/10.48550/arXiv.1909.06146} {{PubMedQA: A Dataset for Biomedical Research Question Answering}}.
\newblock \emph{arXiv e-prints}, page arXiv:1909.06146.

\bibitem[{Kingma and Ba(2014)}]{adam}
Diederik~P. Kingma and Jimmy Ba. 2014.
\newblock \href {https://api.semanticscholar.org/CorpusID:6628106} {Adam: A method for stochastic optimization}.
\newblock \emph{CoRR}, abs/1412.6980.

\bibitem[{{Lewis} et~al.(2020){Lewis}, {Perez}, {Piktus}, {Petroni}, {Karpukhin}, {Goyal}, {K{\"u}ttler}, {Lewis}, {Yih}, {Rockt{\"a}schel}, {Riedel}, and {Kiela}}]{rag}
Patrick {Lewis}, Ethan {Perez}, Aleksandra {Piktus}, Fabio {Petroni}, Vladimir {Karpukhin}, Naman {Goyal}, Heinrich {K{\"u}ttler}, Mike {Lewis}, Wen-tau {Yih}, Tim {Rockt{\"a}schel}, Sebastian {Riedel}, and Douwe {Kiela}. 2020.
\newblock \href {https://doi.org/10.48550/arXiv.2005.11401} {{Retrieval-Augmented Generation for Knowledge-Intensive NLP Tasks}}.
\newblock \emph{arXiv e-prints}, page arXiv:2005.11401.

\bibitem[{Lopes et~al.(2022)Lopes, Kearney, Montero, Hamel, and Brodie}]{kffDebtArticle}
Lunna Lopes, Audrey Kearney, Alex Montero, Liz Hamel, and Mollyann Brodie. 2022.
\newblock \href {https://www.kff.org/report-section/kff-health-care-debt-survey-main-findings/} {Health care debt in the u.s.: The broad consequences of medical and dental bills}.
\newblock Accessed on January 3, 2024.

\bibitem[{Malik et~al.(2021)Malik, Sanjay, Nigam, Ghosh, Guha, Bhattacharya, and Modi}]{court-judgement-prediction}
Vijit Malik, Rishabh Sanjay, Shubham~Kumar Nigam, Kripabandhu Ghosh, Shouvik~Kumar Guha, Arnab Bhattacharya, and Ashutosh Modi. 2021.
\newblock \href {https://doi.org/10.18653/v1/2021.acl-long.313} {{ILDC} for {CJPE}: {I}ndian legal documents corpus for court judgment prediction and explanation}.
\newblock In \emph{Proceedings of the 59th Annual Meeting of the Association for Computational Linguistics and the 11th International Joint Conference on Natural Language Processing (Volume 1: Long Papers)}, pages 4046--4062, Online. Association for Computational Linguistics.

\bibitem[{Medvedeva and Mcbride(2023)}]{legal_judgement_prediction_do_it_right}
Masha Medvedeva and Pauline Mcbride. 2023.
\newblock \href {https://doi.org/10.18653/v1/2023.nllp-1.9} {Legal judgment prediction: If you are going to do it, do it right}.
\newblock In \emph{Proceedings of the Natural Legal Language Processing Workshop 2023}, pages 73--84, Singapore. Association for Computational Linguistics.

\bibitem[{Medvedeva et~al.(2023)Medvedeva, Wieling, and Vols}]{rethinking_automatic_prediction}
Masha Medvedeva, Martijn Wieling, and Michel Vols. 2023.
\newblock \href {https://doi.org/10.1007/s10506-021-09306-3} {Rethinking the field of automatic prediction of court decisions}.
\newblock \emph{Artificial Intelligence and Law}, 31(1):195--212.

\bibitem[{Niklaus et~al.(2021)Niklaus, Chalkidis, and St{\"u}rmer}]{swiss-judgement}
Joel Niklaus, Ilias Chalkidis, and Matthias St{\"u}rmer. 2021.
\newblock \href {https://doi.org/10.18653/v1/2021.nllp-1.3} {{S}wiss-judgment-prediction: A multilingual legal judgment prediction benchmark}.
\newblock In \emph{Proceedings of the Natural Legal Language Processing Workshop 2021}, pages 19--35, Punta Cana, Dominican Republic. Association for Computational Linguistics.

\bibitem[{{Niklaus} and {Giofr{\'e}}(2022)}]{budget-longformer}
Joel {Niklaus} and Daniele {Giofr{\'e}}. 2022.
\newblock \href {https://doi.org/10.48550/arXiv.2211.17135} {{BudgetLongformer: Can we Cheaply Pretrain a SotA Legal Language Model From Scratch?}}
\newblock \emph{arXiv e-prints}, page arXiv:2211.17135.

\bibitem[{{Niklaus} et~al.(2023{\natexlab{a}}){Niklaus}, {Matoshi}, {Rani}, {Galassi}, {St{\"u}rmer}, and {Chalkidis}}]{lextreme}
Joel {Niklaus}, Veton {Matoshi}, Pooja {Rani}, Andrea {Galassi}, Matthias {St{\"u}rmer}, and Ilias {Chalkidis}. 2023{\natexlab{a}}.
\newblock \href {https://doi.org/10.48550/arXiv.2301.13126} {{LEXTREME: A Multi-Lingual and Multi-Task Benchmark for the Legal Domain}}.
\newblock \emph{arXiv e-prints}, page arXiv:2301.13126.

\bibitem[{{Niklaus} et~al.(2023{\natexlab{b}}){Niklaus}, {Matoshi}, {St{\"u}rmer}, {Chalkidis}, and {Ho}}]{multilegalpile}
Joel {Niklaus}, Veton {Matoshi}, Matthias {St{\"u}rmer}, Ilias {Chalkidis}, and Daniel~E. {Ho}. 2023{\natexlab{b}}.
\newblock \href {https://doi.org/10.48550/arXiv.2306.02069} {{MultiLegalPile: A 689GB Multilingual Legal Corpus}}.
\newblock \emph{arXiv e-prints}, page arXiv:2306.02069.

\bibitem[{{Niklaus} et~al.(2024){Niklaus}, {Zheng}, {McCarthy}, {Hahn}, {Rosen}, {Henderson}, {Ho}, {Honke}, {Liang}, and {Manning}}]{flawn-t5}
Joel {Niklaus}, Lucia {Zheng}, Arya~D. {McCarthy}, Christopher {Hahn}, Brian~M. {Rosen}, Peter {Henderson}, Daniel~E. {Ho}, Garrett {Honke}, Percy {Liang}, and Christopher {Manning}. 2024.
\newblock \href {https://doi.org/10.48550/arXiv.2404.02127} {{FLawN-T5: An Empirical Examination of Effective Instruction-Tuning Data Mixtures for Legal Reasoning}}.
\newblock \emph{arXiv e-prints}, page arXiv:2404.02127.

\bibitem[{Obermeyer et~al.(2019)Obermeyer, Powers, Vogeli, and Mullainathan}]{dissecting_racial_bias}
Ziad Obermeyer, Brian Powers, Christine Vogeli, and Sendhil Mullainathan. 2019.
\newblock \href {https://doi.org/10.1126/science.aax2342} {Dissecting racial bias in an algorithm used to manage the health of populations}.
\newblock \emph{Science}, 366(6464):447--453.

\bibitem[{Omiye et~al.(2023)Omiye, Lester, Spichak, Rotemberg, and Daneshjou}]{llms-propagate-race-based-medicine}
Jesutofunmi~A Omiye, Jenna~C Lester, Simon Spichak, Veronica Rotemberg, and Roxana Daneshjou. 2023.
\newblock Large language models propagate race-based medicine.
\newblock \emph{NPJ Digital Medicine}, 6(1):195.

\bibitem[{Pal et~al.(2022)Pal, Umapathi, and Sankarasubbu}]{medmcqa}
Ankit Pal, Logesh~Kumar Umapathi, and Malaikannan Sankarasubbu. 2022.
\newblock \href {https://proceedings.mlr.press/v174/pal22a.html} {Medmcqa: A large-scale multi-subject multi-choice dataset for medical domain question answering}.
\newblock In \emph{Proceedings of the Conference on Health, Inference, and Learning}, volume 174 of \emph{Proceedings of Machine Learning Research}, pages 248--260. PMLR.

\bibitem[{Pollitz et~al.(2023)Pollitz, Pestaina, Lopes, Wallace, and Lo}]{kffDenialSurvey}
Karen Pollitz, Kaye Pestaina, Luna Lopes, Rayna Wallace, and Justin Lo. 2023.
\newblock \href {https://www.kff.org/health-reform/issue-brief/consumer-survey-highlights-problems-with-denied-health-insurance-claims/} {Consumer survey highlights problems with denied health insurance claims}.
\newblock Accessed on January 8, 2024.

\bibitem[{{Rajpurkar} et~al.(2018){Rajpurkar}, {Jia}, and {Liang}}]{squadv2}
Pranav {Rajpurkar}, Robin {Jia}, and Percy {Liang}. 2018.
\newblock \href {https://doi.org/10.48550/arXiv.1806.03822} {{Know What You Don't Know: Unanswerable Questions for SQuAD}}.
\newblock \emph{arXiv e-prints}, page arXiv:1806.03822.

\bibitem[{{Ram} et~al.(2021){Ram}, {Kirstain}, {Berant}, {Globerson}, and {Levy}}]{splinter}
Ori {Ram}, Yuval {Kirstain}, Jonathan {Berant}, Amir {Globerson}, and Omer {Levy}. 2021.
\newblock \href {https://doi.org/10.48550/arXiv.2101.00438} {{Few-Shot Question Answering by Pretraining Span Selection}}.
\newblock \emph{arXiv e-prints}, page arXiv:2101.00438.

\bibitem[{{Ravichander} et~al.(2019){Ravichander}, {Black}, {Wilson}, {Norton}, and {Sadeh}}]{privacyQA}
Abhilasha {Ravichander}, Alan~W {Black}, Shomir {Wilson}, Thomas {Norton}, and Norman {Sadeh}. 2019.
\newblock \href {https://doi.org/10.48550/arXiv.1911.00841} {{Question Answering for Privacy Policies: Combining Computational and Legal Perspectives}}.
\newblock \emph{arXiv e-prints}, page arXiv:1911.00841.

\bibitem[{Ross and Herman(2023)}]{stat-series}
Casey Ross and Bob Herman. 2023.
\newblock \href {https://www.statnews.com/2023/03/13/medicare-advantage-plans-denial-artificial-intelligence/} {Denied by ai: How medicare advantage plans use algorithms to cut off care for seniors in need}.
\newblock \emph{STAT News}.

\bibitem[{{Sanh} et~al.(2019){Sanh}, {Debut}, {Chaumond}, and {Wolf}}]{distilbert}
Victor {Sanh}, Lysandre {Debut}, Julien {Chaumond}, and Thomas {Wolf}. 2019.
\newblock \href {https://doi.org/10.48550/arXiv.1910.01108} {{DistilBERT, a distilled version of BERT: smaller, faster, cheaper and lighter}}.
\newblock \emph{arXiv e-prints}, page arXiv:1910.01108.

\bibitem[{{Singhal} et~al.(2023){Singhal}, {Azizi}, {Tu}, {Mahdavi}, {Wei}, {Chung}, {Scales}, {Tanwani}, {Cole-Lewis}, {Pfohl}, {Payne}, {Seneviratne}, {Gamble}, {Kelly}, {Babiker}, {Sch{\"a}rli}, {Chowdhery}, {Mansfield}, {Demner-Fushman}, {Ag{\"u}era y Arcas}, {Webster}, {Corrado}, {Matias}, {Chou}, {Gottweis}, {Tomasev}, {Liu}, {Rajkomar}, {Barral}, {Semturs}, {Karthikesalingam}, and {Natarajan}}]{deepmindllm}
Karan {Singhal}, Shekoofeh {Azizi}, Tao {Tu}, S.~Sara {Mahdavi}, Jason {Wei}, Hyung~Won {Chung}, Nathan {Scales}, Ajay {Tanwani}, Heather {Cole-Lewis}, Stephen {Pfohl}, Perry {Payne}, Martin {Seneviratne}, Paul {Gamble}, Chris {Kelly}, Abubakr {Babiker}, Nathanael {Sch{\"a}rli}, Aakanksha {Chowdhery}, Philip {Mansfield}, Dina {Demner-Fushman}, Blaise {Ag{\"u}era y Arcas}, Dale {Webster}, Greg~S. {Corrado}, Yossi {Matias}, Katherine {Chou}, Juraj {Gottweis}, Nenad {Tomasev}, Yun {Liu}, Alvin {Rajkomar}, Joelle {Barral}, Christopher {Semturs}, Alan {Karthikesalingam}, and Vivek {Natarajan}. 2023.
\newblock \href {https://doi.org/10.1038/s41586-023-06455-0} {{Publisher Correction: Large language models encode clinical knowledge}}.
\newblock \emph{Nature}, 620(7973):E19--E19.

\bibitem[{{Tafjord} et~al.(2020){Tafjord}, {Dalvi Mishra}, and {Clark}}]{proofwriter}
Oyvind {Tafjord}, Bhavana {Dalvi Mishra}, and Peter {Clark}. 2020.
\newblock \href {https://doi.org/10.48550/arXiv.2012.13048} {{ProofWriter: Generating Implications, Proofs, and Abductive Statements over Natural Language}}.
\newblock \emph{arXiv e-prints}, page arXiv:2012.13048.

\bibitem[{Tuggener et~al.(2020)Tuggener, von D{\"a}niken, Peetz, and Cieliebak}]{ledgar}
Don Tuggener, Pius von D{\"a}niken, Thomas Peetz, and Mark Cieliebak. 2020.
\newblock \href {https://api.semanticscholar.org/CorpusID:218974409} {Ledgar: A large-scale multi-label corpus for text classification of legal provisions in contracts}.
\newblock In \emph{International Conference on Language Resources and Evaluation}.

\bibitem[{{Uddin Ahmad} et~al.(2020){Uddin Ahmad}, {Chi}, {Tian}, and {Chang}}]{policyQA}
Wasi {Uddin Ahmad}, Jianfeng {Chi}, Yuan {Tian}, and Kai-Wei {Chang}. 2020.
\newblock \href {https://doi.org/10.48550/arXiv.2010.02557} {{PolicyQA: A Reading Comprehension Dataset for Privacy Policies}}.
\newblock \emph{arXiv e-prints}, page arXiv:2010.02557.

\bibitem[{{Wang} et~al.(2023){Wang}, {Scardigli}, {Tang}, {Chen}, {Levkin}, {Chen}, {Ball}, {Woodside}, {Zhang}, and {Hendrycks}}]{maud}
Steven~H. {Wang}, Antoine {Scardigli}, Leonard {Tang}, Wei {Chen}, Dimitry {Levkin}, Anya {Chen}, Spencer {Ball}, Thomas {Woodside}, Oliver {Zhang}, and Dan {Hendrycks}. 2023.
\newblock \href {https://doi.org/10.48550/arXiv.2301.00876} {{MAUD: An Expert-Annotated Legal NLP Dataset for Merger Agreement Understanding}}.
\newblock \emph{arXiv e-prints}, page arXiv:2301.00876.

\bibitem[{{Yu} et~al.(2022){Yu}, {Sun}, {Xu}, {Dong}, {Chen}, {Xu}, and {Wen}}]{legal-case-matching}
Weijie {Yu}, Zhongxiang {Sun}, Jun {Xu}, Zhenhua {Dong}, Xu~{Chen}, Hongteng {Xu}, and Ji-Rong {Wen}. 2022.
\newblock \href {https://doi.org/10.48550/arXiv.2207.04182} {{Explainable Legal Case Matching via Inverse Optimal Transport-based Rationale Extraction}}.
\newblock \emph{arXiv e-prints}, page arXiv:2207.04182.

\bibitem[{{Zheng} et~al.(2021{\natexlab{a}}){Zheng}, {Guha}, {Anderson}, {Henderson}, and {Ho}}]{when-does-pretraining-help}
Lucia {Zheng}, Neel {Guha}, Brandon~R. {Anderson}, Peter {Henderson}, and Daniel~E. {Ho}. 2021{\natexlab{a}}.
\newblock \href {https://doi.org/10.48550/arXiv.2104.08671} {{When Does Pretraining Help? Assessing Self-Supervised Learning for Law and the CaseHOLD Dataset}}.
\newblock \emph{arXiv e-prints}, page arXiv:2104.08671.

\bibitem[{{Zheng} et~al.(2021{\natexlab{b}}){Zheng}, {Guha}, {Anderson}, {Henderson}, and {Ho}}]{casehold}
Lucia {Zheng}, Neel {Guha}, Brandon~R. {Anderson}, Peter {Henderson}, and Daniel~E. {Ho}. 2021{\natexlab{b}}.
\newblock \href {https://doi.org/10.48550/arXiv.2104.08671} {{When Does Pretraining Help? Assessing Self-Supervised Learning for Law and the CaseHOLD Dataset}}.
\newblock \emph{arXiv e-prints}, page arXiv:2104.08671.

\end{thebibliography}

\appendix

\section{Detailed Dataset Composition}
\label{sec:detailedcomposition}

We now describe the data in each of the categories that partition our dataset. For each document we made an effort to determine relevant usage restrictions. We also document the results of those search efforts in this section.

    \textbf{Legal.}
    \begin{itemize}
    \item State and Territory Code and Statutes.

    We scraped the full text of statutes for all U.S. states and territories from \url{https://justia.com}. In each case, we incorporated only the text from the most recent yearly release, to avoid large-scale redundancy. This text is public domain.

    \item U.S. Code.
    
    We scraped the full text of U.S. code from \url{https://uscode.house.gov/} (release point 118-30). This text is public domain.

    \item U.S. Code of Federal Regulations.

    We scraped the full text of the electronic code of federal regulations from \url{https://www.govinfo.gov/bulkdata/ECFR}. This text is public domain.
    
    \item U.S. Federal Register.

    We scraped the full text of the Federal Register from \url{https://www.govinfo.gov/bulkdata/xml/FR}. This text is public domain.
    \end{itemize}
    
    \textbf{Regulatory Guidance.}

    \begin{itemize}
    \item General Regulatory Guidance from CMS.
    
    We scraped the \href{https://www.cms.gov/medicare/regulations-guidance/manuals/internet-only-manuals-ioms}{internet only regulatory guidance manuals} from the Centers for Medicare and Medicaid Services (CMS). \href{https://www.cms.gov/medicare/regulations-guidance/manuals/internet-only-manuals-ioms}{According to CMS}, this text is comprised of ``issuances, day-to-day operating instructions, policies, and procedures that are based on statutes, regulations, guidelines, models, and directives". This text is public domain.

    \item Federal Policy Guidance on Medicaid from CMS.

    CMS releases \href{https://www.medicaid.gov/federal-policy-guidance/index.html}{guidance} on Medicaid program implementation and compliance. We scraped all of the historical guidance. This text is public domain.
    
    \end{itemize}

\textbf{Coverage Rules, Contracts, and Medical Policies.}

    \begin{itemize}
    \item Medicare LCDs and NCDs.

    Coverage rules in Medicare are informed by \href{https://www.cms.gov/medicare/coverage/determination-process}{National Coverage Determinations} (NCD) and \href{https://www.cms.gov/medicare/coverage/determination-process/local}{Local Coverage Determinations} (LCD) that change
    over time. We scraped NCDs and LCDs active as of early 2024. This text is public domain.

    \item State Medicaid Handbooks.

    States release handbooks detailing benefit structures for their Medicaid programs (see e.g. the 2024 MaineCare \href{https://www.maine.gov/dhhs/sites/maine.gov.dhhs/files/inline-files/mainecare-member-handbook.pdf}{handbook}). We scraped some such handbooks. This text is public domain.

    \end{itemize}

\textbf{Opinion, Policy, and Summary.}

\begin{itemize}
\item HHS OIG Portfolio Reports.

    The Health and Human Services Office of the Inspector General (HHS OIG) releases reports detailing agency recommendations. We scraped reports focused on recommendations to improve regulatory compliance. This text is public domain.

\item HHS OIG Congressional Testimony.

    HHS OIG releases \href{https://oig.hhs.gov/newsroom/testimony/}{testimony} from hearings that their staff take part in. We scraped all historical testimony for which there are transcripts. This text is public domain.

\item HHS OIG Red Books.

    The HHS OIG historically released `red books' concerned with program cost saving recommendations. We scraped the historical catalog. This text is public domain.

\item HHS OIG Top Management and Performance Challenges Facing HHS Reports. 

    HHS OIG has a statutory obligation to release \href{https://oig.hhs.gov/reports/all/?report-type=TMC}{annual reports} detailing their biggest challenges. We scraped the historical catalog. This text is public domain.

\item Medicaid Integrity Reports.

The HHS OIG historically released \href{https://www.cms.gov/medicare/medicaid-coordination/center-program-integrity/reports-guidance}{annual reports} about the status of the integrity of the Medicaid program.  We scraped the historical catalog. This text is public domain.

\item HHS OIG Semiannual Reports to Congress.

    The HHS OIG delivers \href{https://oig.hhs.gov/reports/all/?report-type=SAR}{semiannual reports} to congress detailing their activities. We scraped the historical catalog. This text is public domain.

\item HHS OIG Healthcare Fraud and Abuse Control Program Reports.

    The HHS OIG provides \href{https://oig.hhs.gov/reports/all/?report-type=HCFAC}{reports} on the funding and activities of the Healthcare Fraud and Abuse Control Program, which is aimed at preventing and regulating fraud. We scraped the historical catalog. The text is public domain.

\item HHS OEI Reports.

    The HHS \href{https://oig.hhs.gov/about-oig/office-evaluation-and-inspections/}{Office of Evaluations and Inspections} (OEI) ``conducts national evaluations of HHS programs from a broad, issue-based perspective". We scraped a historical catalog of their reports. The text is public domain.

\end{itemize}

\textbf{Case Descriptions.}
    \begin{itemize}
    \item CA CDI External Appeals.

        The California Department of Insurance (CDI) manages an Independent Medical Review program, and maintains a \href{https://interactive.web.insurance.ca.gov/apex_extprd/f?p=192:1:5713079812566:::::}{database} of outcomes for appeal cases from the program. We scraped summaries from those cases. The text is public domain.
        
    \item CA DMHC External Appeals.

        The California Department of Managed Healthcare (DMHC) manages a separate Independent Medical Review program, and maintains a \href{https://data.chhs.ca.gov/dataset/independent-medical-review-imr-determinations-trend}{database} of its outcomes. We scraped summaries for those cases. The text is public domain.
        
    \item Medicare Part C and D Appeals.

        CMS maintains a database with Level 2 Medicare appeals from Part C and Part D. We scraped \href{https://www.cms.gov/medicare/appeals-grievances/appeals-decision-search-part-c-d?planType=Part+C&sort=desc}{summaries} for those cases. The text is public domain.

    \item New York DFS External Appeals.

        The New York Department of Financial Services (DFS) manages an external appeal program, and maintains a \href{https://www.dfs.ny.gov/public-appeal/search}{database} of its outcomes. We scraped summaries for those cases. The text is public domain.
    \end{itemize}

\textbf{Clinical Guidelines.}
    \begin{itemize}
    \item FDA Guideline Reports.

        The Food and Drug Administration (FDA) \href{https://www.fda.gov/regulatory-information/search-fda-guidance-documents#guidancesearch}{releases} official Guidance Documents on a variety of topics. We scraped all such official guidance. The text is public domain.
        
    \item Meditron Clinical Guidelines Dataset.

    In \cite{meditron}, the authors introduced a dataset of clinical medical guidelines
    from a diverse collection of permissively licensed sources. We included  a part of this dataset in our corpus. According to that work, the subset of the text we incorporated allows for redistribution.

    \end{itemize}

\section{Source Data Usage Restrictions}
\label{usage_restrictions}

We made an effort to only use data which is permissively licensed and allows for redistribution. However, it is possible we made a mistake. We encourage those who use our data to ensure they are also following license terms for our source data, and to contact us directly if they believe we are not following such terms. 

In some contexts we were unable to determine or verify usage restrictions for documents we expect are public domain, based on their source. In other cases, the usage terms for documents of interest explicitly prohibited redistribution. We did not include either type of document in our corpus, but some are documented in our scraper code.

\section{Appeal Adjudication Task Formulation}
\label{sec:task-description}

Our task involves \emph{external} appeals because they are a proxy for unbiased adjudication. External appeal processes vary, but a prototypical example is that of Affordable Care Act marketplaces. Companies serve as third party reviewers for appeals, and assess the merit of denials. Their determinations are then binding.

There are two distinct tasks one might assume we are considering here. One is to predict the likelihood that a case merits overturn (according to e.g. law, and contracts). Another is to predict the likelihood that a case would be overturned, were it appealed (on the basis of e.g. law, contracts, individual opinion, bias, perverse monetary incentives, etc.).

Both of these tasks are meaningful, and could improve systems and patient outcomes. Here we are modeling the latter task, rather than the former. This distinction has many important implications. Most notable is the fact that reviewer adjudication may in practice encode biases, be unjust, or simply make mistakes. For this reason what \emph{ought to} happen need not be the same as what \emph{is likely} to happen in an external appeal case. Because we model the latter task, our models encode historical bias. See Section \ref{discussion} for further discussion.

\section{Annotation Guidelines}
\label{sec:annotation-guidelines}

Our annotation guidelines are an evolving work in progress, and future guidelines we provide to annotators will be released along with future versions of our dataset. Here we share the current iteration of the high level
guidance for the background span annotation and sufficiency labeling subtasks. Our annotation instructions contain additional detailed guidance not included here.

\begin{displayquote}
\textbf{Background Span Selection}

This is the maximal subset of the Case Summary which could reasonably be known by a patient or caseworker at the time of the appeal submission. This includes, but is not necessarily limited to, all details provided that describe the situation leading up to the adverse coverage denial, and all actions taken by the insurer before an appeal was submitted to an independent reviewer. This should not include information which directly or indirectly indicates the adjudication outcome, which constitutes the opinion of the reviewer, or which could not be known prior to appeal submission.

The background context span labeling task is particularly subtle, as there are often summary descriptions of background context made by a reviewer imbued with potentially leaking, opinion based, or subjective language. When in doubt about whether a particular subset of a case summary leaks the adjudication rationale, viewpoint, or decision of a reviewer, do \textbf{not} highlight the subset.

\textbf{Background Span Sufficiency Scoring}

\underline{General Guidance}

Rate the selected background context span on a scale from 1 to 4, to indicate the degree to which the background context alone is sufficient to make an informed guess about the likelihood of denial overturn. Lower scores indicate relative insufficiency of the context to make such an informed prediction, and higher scores indicate relative sufficiency.

\end{displayquote}
Table \ref{sufficiency_guidelines} shows a specific rubric we're actively developing to support annotators in making sufficiency determinations.

\begin{table*}[h]
    \centering
    \begin{tabular}{l|p{20em}|p{12em}}
    \hline
    \textbf{Score} & \textbf{Criteria} & \textbf{Example Selection} \\
    \hline
    1 & Little to no specific context for the denial is provided. 
    \vspace{.5em}

    Neither a specific condition, 
    specific service or treatment, nor specific drug is stated in the description. & 
    "An enrollee has requested reimbursement for laboratory testing (genetic or chemical)." \\
    \hline
    2 & Some specific context for the denial is provided. 
    \vspace{.5em}

    At minimum, the description mentions either the specific condition / medical situation 
    being treated, or the specific service / treatment. & "An enrollee has requested bilateral sacroiliac (SI) joint fusion surgery for treatment of her medical condition."  \\
    \hline
    3 & Some specific context for the denial is provided.
    \vspace{.5em}

    At minimum, the description mentions the specific condition / medical situation being treated, and the specific service / treatment.  & "The parent of a male enrollee has requested authorization and coverage for immunotherapy with nivolumab (Opdivo). The Health Plan has denied this request indicating that the requested medication is considered investigational for treatment of the enrollees grade IV glioblastoma." \\
\hline
    4 & Context for the denial additional to the condition / medical event and service / treatment is provided. 
    \vspace{.5em}
    
    At minimum the context includes the condition / medical situation being treated, the service / treatment, and some description of the specific events or rationale leading to the coverage request.

    & "A 62-year-old female enrollee has requested reimbursement for substance abuse residential rehabilitation services provided from 2/17/14 through 3/29/14. The Health Insurer has denied this request indicating that the services at issue
    were not medically necessary for treatment of the enrollees alcohol dependence and major depressive disorder. This
    patient had multiple medical problems including thyroid disease and chronic pain. She also had depression
    and had previously made a suicide attempt that prompted a one-to-one attendant until 3/3/14. During treatment,
    the patient received care for her alcohol dependence and major depressive disorder." \\
\hline
    \end{tabular}
    \caption{\label{sufficiency_guidelines} Sufficiency score annotation guidelines. }
    \end{table*}

\section{Experimental Methodology}
\label{sec:experimental-setup}

We now describe more details about our modeling and experimental setup. All experimental code has released under a permissive license, and we refer those interested in additional specific details to the code.

\textbf{Background Span Selector Training}
We trained a model to extract background spans from each of the raw case descriptions used for our benchmark. To do this we finetuned DistilBERT on token classification, using 1,000 manual span annotations for the combined training and test set. We initially experimented with multiple span-selection architectures, unsupervised bootstrapping approaches such as Splinter \cite{splinter}, and hosted language models. We found the best initial results by casting span selection as an equivalent token classification problem, perhaps due to the small size of our manually annotated training set. We trained using Adam \cite{adam}, with a learning rate of $1 \times 10^{-5}$.

\textbf{Sufficiency Model Training}
We trained a model to produce binary `sufficiency' pseudo-labels for each of the raw case descriptions used for our benchmark. To do this we finetuned DistilBERT \cite{distilbert} on sequence classification, using 1,000 manual sufficiency annotations for the combined training and test set. We trained using Adam, with a learning rate of $2 \times 10^{-8}$.

\textbf{Outcome Prediction Experiments}

We finetuned and evaluated three pretrained bidirectional transformers with different sizes and pretraining corpora. We finetuned the models on our task by mounting a classification head, and freely training the resulting classifiers on our training set. In each case, we trained using Adam, with a learning rate of $2 \times 10^{-6}$.

Finally, we evaluated gpt-4o-mini-2024-07-18 on our benchmark in a two-shot setting. We included a detailed system prompt explaining the task, which is being released with our code.

\section{Case Adjudication Dataset Examples}
\label{sec:case-adjudication-dataset}

We provide a few examples of our pseudo-annotated case background summaries in Figures \ref{Case Example 1}, \ref{Case Example 2}, \ref{Case Example 3}, and \ref{Case Example 4}. 

We selected one example for each (sufficiency label, outcome) category. For each example
we provide the source text, with the selected (possibly disconnected) spans highlighted in red. The span selection and sufficiency label are produced by our Distilbert variants. We also show the concatenation of the selected spans, labeled as ``Extracted Case Background'' for readability. This concatenation is what is used as the background summary in our task benchmark.

\newpage

\begin{figure*}[ht]

\begin{tcolorbox}[colback=blue!3!white, colframe=blue!35!black, title=Source Text, enhanced, breakable]
Summary Reviewer 2 \textcolor{BrickRed}{A 61-year-old male enrollee has requested authorization and coverage for proton beam radiation therapy. The Health Insurer has denied this request indicating that the requested services are considered investigational for treatment of the enrollees colorectal cancer with liver metastases.} The physician reviewer found that there is no medical literature to suggest that treatment with proton beam radiation offers a clinical benefit compared to conventional radiation therapy techniques for rectal cancer (Allen, et al). There is no data submitted that suggests that treatment with proton techniques will be able to improve sparing of the small bowel compared to standard intensity modulated radiation or three-dimensional (3D) conformal techniques with photons. The National Comprehensive Cancer Network (NCCN) guidelines do not support the use of proton therapy in this clinical setting. In the majority of cases, treatment with photon techniques is advised, especially in the palliative setting. Moreover, there is a lack of level I evidence to support the use of proton therapy for the treatment of rectal cancer. Based on these reasons, the requested proton beam therapy is not likely to be more beneficial for treatment of the patients medical condition than any available standard therapy. Based upon the information set forth above, the requested services are not likely to be more beneficial for treatment of the patients medical condition than any available standard therapy. The Health Insurers denial should be upheld.
\end{tcolorbox}

            \begin{tcolorbox}[colback=green!3!white, colframe=green!35!black, title=Extracted Case Background, enhanced, breakable]
                A 61-year-old male enrollee has requested authorization and coverage for proton beam radiation therapy. The Health Insurer has denied this request indicating that the requested services are considered investigational for treatment of the enrollees colorectal cancer with liver metastases.
            \end{tcolorbox}

    \begin{tcolorbox}[colback=green!3!white, colframe=green!35!black, title=Outcome, enhanced, breakable]
    \textbf{Upheld}
    \end{tcolorbox}

    \begin{tcolorbox}[colback=green!3!white, colframe=green!35!black, title=Sufficiency Label, enhanced, breakable]
    \textbf{Sufficient}
    \end{tcolorbox}

\caption{A case summary corresponding to an appeal that was upheld. Background context extracted by our model appears in red. Our sufficiency model labeled this background context as `Sufficient' for making an informed prediction about the expected case outcome.}
\label{Case Example 1}
\end{figure*}
\newpage

\begin{figure*}[ht]

\begin{tcolorbox}[colback=blue!3!white, colframe=blue!35!black, title=Source Text, enhanced, breakable]
    Summary Reviewer 2 \textcolor{BrickRed}{A 70-year-old male enrollee has requested reimbursement for service code 97139, unlisted therapeutic procedure, performed on 4/20/20. The Health Insurer has denied this request and reported that the services at issue were investigational for the treatment of the enrolleeas medical condition.} The physician reviewer found that the submitted documentation fails to demonstrate the superior efficacy of the services at issue. The services at issue were not likely to have been superior over standard therapy of physical therapy, epidural steroid injections, modification of the activity that may exacerbate the pain, and medication management. A review of the literature does not reveal sufficient evidence regarding the effectiveness of this procedure for this patientas condition. Therefore, the unlisted therapeutic procedure provided on 4/20/20 was not likely to have been more beneficial than any available standard therapy. 
    \end{tcolorbox}

        \begin{tcolorbox}[colback=green!3!white, colframe=green!35!black, title=Extracted Case Background, enhanced, breakable]
            A 70-year-old male enrollee has requested reimbursement for service code 97139, unlisted therapeutic procedure, performed on 4/20/20. The Health Insurer has denied this request and reported that the services at issue were investigational for the treatment of the enrolleeas medical condition. 
    \end{tcolorbox}

    \begin{tcolorbox}[colback=green!3!white, colframe=green!35!black, title=Outcome, enhanced, breakable]
    \textbf{Upheld}
    \end{tcolorbox}

    \begin{tcolorbox}[colback=green!3!white, colframe=green!35!black, title=Sufficiency Label, enhanced, breakable]
    \textbf{Insufficient}
    \end{tcolorbox}
\caption{A case summary corresponding to an appeal that was upheld. 
Background context extracted by our model appears in red. 
Our sufficiency model labeled this background context as 'Insufficient' for making an informed prediction 
about the expected case outcome.}
\label{Case Example 2}
\end{figure*}
\newpage

\begin{figure*}[ht]
    \begin{tcolorbox}[colback=blue!3!white, colframe=blue!35!black, title=Source Text, enhanced, breakable]
    Nature of Statutory Criteria/Case Summary:  \textcolor{BrickRed}{An enrollee has requested authorization and coverage for Octagam (IVIG).F}indings: Two out of three physician reviewers found that Octagam (IVIG) is likely to be more beneficial for you than any available standard therapy.AIDP is a known complication following the administration of several vaccines as well as a complication following illness from several pathogens. CIDP is the chronic form of AIDP. AIDP and CIDP have both been documented in the literature after COVID-19 vaccinations as well as COVID-19 illness. Evidence for intravenous immune globulin (IVIG) in the setting of AIDP/CIDP, specifically secondary to COVID infection/vaccination related issues, is limited. AIDP and CIDP are well-known entities with established treatment recommendations. The management of these diseases involves the administration of IVIG. Additionally, IVIG is approved by the U.S. Food and Drug Administration in the setting of CIDP. There is no qualification on the pathogen or vaccine responsible to bring about the pathological immune response. Additionally, considering the novelty of COVID-19, it can be argued a standard therapy does not exist. In this case, neurophysiological testing is consistent with a severe sensory polyneuropathy in the lower extremities. Therefore, Octagam (IVIG) is likely to be more beneficial than any other available standard treatments for the patient's medical condition.Final Result: The reviewers determined that the requested medication is likely to be more beneficial than any available standard treatment for the patient's medical condition. Therefore, the Health Plan's denial should be overturned.Credentials/Qualifications: The reviewer 1, reviewer 2, and reviewer 3 are board certified in neurology. All reviewers are actively practicing. The reviewers are experts in the treatment of the enrollee's medical condition and knowledgeable about the proposed treatment through recent or current actual clinical experience treating those with the same or a similar medical condition.
    \end{tcolorbox}

    \begin{tcolorbox}[colback=green!3!white, colframe=green!35!black, title=Extracted Case Background, enhanced, breakable]
    An enrollee has requested authorization and coverage for Octagam (IVIG).
    \end{tcolorbox}

    \begin{tcolorbox}[colback=green!3!white, colframe=green!35!black, title=Outcome, enhanced, breakable]
    \textbf{Overturned}
    \end{tcolorbox}

    \begin{tcolorbox}[colback=green!3!white, colframe=green!35!black, title=Sufficiency Label, enhanced, breakable]
    \textbf{Insufficient}
    \end{tcolorbox}

\caption{A case summary corresponding to an appeal that was overturned. Background context extracted by our model appears in red. Our sufficiency model labeled this background context as 'Insufficient' for making an informed prediction about the expected case outcome. In this case, the background context merely states a request for coverage of a medication, without any further details about the request.}
\label{Case Example 3}
\end{figure*}

\begin{figure*}[ht]
    \begin{tcolorbox}[colback=blue!3!white, colframe=blue!35!black, title=Source Text, enhanced, breakable]
    Summary Reviewer \textcolor{BrickRed}{62-year-old female enrollee has requested reimbursement for substance abuse residential rehabilitation services provided from 2/17/14 through 3/29/14. The Health Insurer has denied this request indicating that the services at issue were not medically necessary for treatment of the enrollees alcohol dependence and major depressive disorder.} 
    The physician reviewer found that due to multiple factors and unique considerations, determination of the 
    appropriate level of care for patients with addictive disorders complicated by depressive illness and medical 
    comorbidity is a challenging endeavor. Assisting with this process, the American Psychiatric Association (APA) 
    guidelines state that residential treatment is indicated primarily for individuals who do not meet clinical criteria 
    for hospitalization but whose lives and social interactions have come to focus exclusively on substance use and who 
    currently lack sufficient motivation and/or substance-free social supports to remain abstinent in an ambulatory setting. 
    \textcolor{BrickRed}{This patient had multiple medical problems including thyroid disease and chronic pain. She also had significant depression and had previously made a serious suicide attempt that prompted a one-to-one attendant until 3/3/14. During treatment, the patient received clinically appropriate care for her alcohol dependence and major depressive disorder.} 
    The comprehensive services were delivered in accordance with an individual treatment plan. Further, the services were reasonably expected to improve the patients condition and prevent a more serious episode of illness. The placement also provided opportunity for marital harmonization as her spouse attended the family program. Considering the clinical presentation and perusal of the peer-reviewed literature, residential placement with access to specialty services 24-hours a day was a safe, appropriate disposition consistent with good medical practice. Accordingly, the services at issue were medically necessary for treatment of the patients medical condition. In conclusion, the services at issue were medically necessary for treatment of the patients medical condition. The Health Insurers denial should be overturned. 
    \end{tcolorbox}

    \begin{tcolorbox}[colback=green!3!white, colframe=green!35!black, title=Extracted Case Background, enhanced, breakable]
        A 62-year-old female enrollee has requested reimbursement for substance abuse residential rehabilitation services provided from 2/17/14 through 3/29/14. 
        The Health Insurer has denied this request indicating that the services at issue were not medically necessary for treatment of the enrollees alcohol dependence and major depressive disorder. 
        This patient had multiple medical problems including thyroid disease and chronic pain. 
        She also had significant depression and had previously made a serious suicide attempt that prompted a one-to-one attendant until 3/3/14. 
        During treatment, the patient received clinically appropriate care for her alcohol dependence and major depressive disorder. 
    \end{tcolorbox}

    \begin{tcolorbox}[colback=green!3!white, colframe=green!35!black, title=Outcome, enhanced, breakable]
    \textbf{Overturned}
    \end{tcolorbox}

    \begin{tcolorbox}[colback=green!3!white, colframe=green!35!black, title=Sufficiency Label, enhanced, breakable]
    \textbf{Sufficient}
    \end{tcolorbox}

\caption{A case summary corresponding to an appeal that was overturned. Background context extracted by our model appears in red. Our sufficiency model labeled this background context as `Sufficient' for making an informed prediction about the expected case outcome. 
In this case, the selected background context is not contiguous. 
It also appears our background selection model inappropriately selected some subspans that questionably leak the reviewer's expert opinion. 
Namely, the inclusion of `significant', `serious', and `clinically appropriate' in the sentences: ``She also had significant depression and had previously made a serious suicide attempt that prompted a one-to-one attendant until 3/3/14. During treatment, the patient received clinically appropriate care for her alcohol dependence and major depressive disorder.''. The intent of the more detailed annotation guidelines (not shown in this text) in this case would be to instruct 
an annotator to not include those qualifying phrases.}
\label{Case Example 4}
\end{figure*}

\end{document}